\documentclass[apj,graphicx]{emulateapj}
\usepackage{apjfonts}

\usepackage{graphicx, natbib}
\def\Swift{{\em Swift}}

\def\XMM{{\em XMM-Newton}}

\begin{document}

\shorttitle{V404 Cyg and Cen X-4}
\shortauthors{Bernardini et al.}

\title{On the Optical -- X-ray correlation from outburst to quiescence in Low Mass X-ray Binaries: the representative cases of V404 Cyg and Cen X-4}

\author{F.~Bernardini \altaffilmark{1}}
\author{D.M.~Russell\altaffilmark{1}}
\author{K.~Koljonen\altaffilmark{1}}
\author{L.~Stella\altaffilmark{2}}
\author{R.I.~Hynes\altaffilmark{3}}
\author{S.~Corbel\altaffilmark{4}}
\email{bernardini@nyu.edu}

\affil{\altaffilmark{1}{New York University Abu Dhabi, P.O. Box 129188, Abu Dhabi, United Arab Emirates;bernardini@nyu.edu}}
\affil{\altaffilmark{2}{INAF - Osservatorio astronomico di Roma, Via Frascati 44, Monteporzio I-00040, Catone (Roma), Italy}}
\affil{\altaffilmark{3}{Department of Physics and Astronomy, Louisiana State University, 202 Nicholson Hall, Tower Drive, Baton Rouge, LA 70803, USA}}
\affil{\altaffilmark{4}{Laboratoire AIM (CEA/IRFU - CNRS/INSU - Universit\'e Paris Diderot), CEA DSM/IRFU/SAp, F-91191 Gif-sur-Yvette, France}}

\begin{abstract}

Low mass X-ray binaries (LMXBs) show evidence of a global correlation of debated origin between X-ray and optical luminosity. We study for the first time this correlation in two transient LMXBs, the black hole V404 Cyg and the neutron star Cen X-4, over 6 orders of magnitude in X-ray luminosity, from outburst to quiescence. After subtracting the contribution from the companion star, the Cen X-4 data can be described by a single power law correlation of the form $L_{opt}\propto\,L_{X}^{0.44}$, consistent with disk reprocessing. We find a similar correlation slope for V404 Cyg in quiescence (0.46) and a steeper one (0.56) in the outburst hard state of 1989. However, V404 Cyg is about 160--280 times optically brighter, at a given 3--9 keV X-ray luminosity, compared to Cen X-4. This ratio is a factor of 10 smaller in quiescence, where the normalization of the V404 Cyg correlation also changes. Once the bolometric X-ray emission is considered and the known main differences between V404 Cyg and Cen X-4 are taken into account (a larger compact object mass, accretion disk size, and the presence of a strong jet contribution in the hard state for the black hole system) the two systems lie on the same correlation. In V404 Cyg, the jet dominates spectrally at optical-infrared frequencies during the hard state, but makes a negligible contribution in quiescence, which may account for the change in its correlation slope and normalization. These results provide a benchmark to compare with data from the 2015 outburst of V404 Cyg and, potentially, other transient LMXBs as well.

\end{abstract}

\keywords{accretion, accretion disks --- black hole physics --- stars: black holes: individual (V404 Cyg) --- X-rays: binaries, stars: neutron: individual (Cen X-4)}

\section{Introduction}

Low Mass X-ray binaries (LMXBs) are binary systems in which a compact object, a black hole (BH) or a neutron star (NS), accretes matter from a low mass companion. Many LMXBs are transient and alternate long periods of quiescence, lasting up to many decades, where the X-ray luminosity is low (of the order of few $10^{30}$ -- few $ 10^{33}$ erg/s), with shorter periods of activity called outbursts, lasting weeks to years, where the X-ray emission is much higher (up to $10^{37}-10^{38}$ erg/s) and may approach the Eddington limit. The characteristics of the quiescence/outburst cycle are broadly explained by the disk instability model \citep[DIM, see e.g.][]{cannizzo93,lasota01}. Matter from the companion star accumulates in the accretion disk during quiescence, when the X-ray emission is low; then, during outburst, it is accreted towards the compact object and the X-ray luminosity increases by orders of magnitude. 

At the beginning and at the end of an outburst, LMXBs are found in the so called low-hard state \citep[see e.g.][for a review on XB spectral states]{belloni10}, where the X-ray spectrum is dominated by a non-thermal power law. The origin of this spectral component, which extends up to the hard X-ray frequencies, is still unclear. It is commonly accepted that it is produced in a hot corona that surrounds the compact object and Compton up-scatters the seed photons emitted by the accretion disk (in the UV and soft X-ray frequencies), or at the base of a compact jet \citep[see][and references therein]{markoff05}.

In black hole X-ray binaries (BHXBs) the quiescent and low-hard states share similar properties. For example, a steady compact jet is detected in both states and produces a flat or slightly inverted radio spectrum. Quiescence has been long seen as a low-luminosity version of the hard state. However, recent observations suggest that they could be two distinct states. For example, in some cases the BHXB spectra soften toward quiescence, as a result of the steepening of the X-ray power law \citep{plotkin15}. 

The emission processes in the optical-infrared (OIR) and UV bands are still poorly known for LMXBs. Several different, and possibly competing, emission mechanisms are expected to contribute at the relevant frequencies, depending on the source type and spectral state. The spectral energy distribution (SED) of quiescent LMXBs is characterized by an excess in the OIR (up to the UV band), with respect to the emission of the companion star alone. This excess is expected to trace the accretion flow; however, the physical mechanisms producing it and the exact location in the binary where these processes take place are still debated. 
Optical studies suggest that the excess could be intrinsic emission from a hot inner region of the disk \citep{hynes12}, perhaps due to magnetic reconnections \citep[see e.g.][]{hynes03,zurita03}, or  emission produced by the gravitational energy released by matter close to the circularization radius in BH systems, and by the interaction of the pulsar relativistic wind and the inflowing matter in NS systems \citep{cam&stell}. It could also be produced by X-ray reprocessing from an irradiated disk \citep[][]{cackett13,bernardini13} or by compact jets whose emission may also extend up to X-ray and $\gamma$-ray energies \citep[see e.g.][]{markoff05,nowak11,peer12,zdziarski14}.

In transient LMXBs a fruitful method to understand the origin of the emission at low frequencies (radio, OIR and UV) is to study its correlation with the X-ray emission using simultaneous observations. 
When comparing radio and X-ray fluxes, a tight power law correlation was found \citep[$L_{R}\propto L^{0.6-0.7}_{X}$;][]{corbel03,gallo03,gallo06,corbel08} suggesting a strong link between radio and X-ray emission, and that the X-ray band could be dominated by synchrotron emission from the jet \citep[][]{markoff01,markoff03}. Later on, a number of outliers departing from the correlation were identified, showing that two independent tracks are present \citep[with slope 0.68 and 0.98;][]{gallo12}. The origin of these two tracks is still unclear. A similar study was also conducted on NS LMXBs and a radio--X-ray correlation was found with a steeper slope of 1.4. NS are fainter in the radio, for a given X-ray luminosity, compared to BHs \citep{migliari06}. 

The spectral signature of a synchrotron emitting jet has been found in several BHXBs during quiescence (including V404 Cyg), either at radio \citep{gallo05,gallo06,gallo14,plotkin15} or infrared \citep{gallo07,gelino10,shahbaz13} frequencies. Although infrared excesses in the broadband spectra have been also interpreted as due to dust or circumbinary emission \citep{muno06}, in some cases at least this emission is variable, implying a synchrotron origin \citep{russell13,shahbaz13}. NSXB jets, if they exist in quiescence, are expected to be much fainter than BHXB jets \citep{migliari06}, and have only been reported in one quiescent system \citep{baglio13} and one source at a low (but not quiescent) luminosity \citep{deller14}.

OIR and X-ray studies, including data from the NS system Cen X-4 and the BH system V404 Cyg, have shown that there exists a global correlation with a slope $0.6\pm0.1$ in NSXBs in outburst and BHXBs in the outburst hard state, also extending to quiescence \citep{russell06,russell07}. This suggests that synchrotron emission from the jet can play an important role also at IR and optical frequencies. However, X-ray reprocessing from the disk \citep[see also][]{vanpar94} and intrinsic emission from a viscously heated disk can also account for the observed correlation, depending on frequency band (optical, NIR) and source type. The exact contribution from different emission processes depends on a number of parameters, including the disk size and the shape of the jet spectrum \citep{russell06}.

In the case of the BH system GX 339-4 in the hard state the optical emission shows a power law correlation with a slope of $0.44\pm0.01$ \citep{coriat09}, while 
the NIR emission correlates with the X-ray following a broken power law correlation. These authors conclude that the NIR is jet-dominated, while in the optical another component dominates, probably reprocessing in the disk \citep[see also][]{homan05}, and the jet contributes about $40\%$.
In BH system XTE J1817-330 the near-UV flux and the hard X-ray flux show a power law correlation with index 0.5 in the decline from its 2006 outburst up to the return to the low/hard state, suggesting that X-ray reprocessing is the dominating mechanism at near-UV frequency \citep{rykoff07}.

More recent studies show that irradiation and reprocessing play an important role also in quiescence. 
In the case of the NS Cen X-4, when strictly simultaneous optical/UV and X-ray data were used, a short-timescale ($\lesssim$days) correlation was found. The shape of this is correlation is a power law with slope close to 0.5 and it was interpreted as X-ray reprocessing from the accretion disk and another extended surface, presumably the companion star \citep[]{bernardini13,cackett13}. \cite{hynes04} showed that (see their Fig. 1 and 2) the H$_{\alpha}$ emission in the BH V404 Cyg in quiescence is positively correlated with the X-ray flux on very short timescales (less than hours, perhaps hundreds of seconds). Moreover, the H$_{\alpha}$ line is double-peaked implying that it is produced in the disc. The combination of the two results suggests that the excess optical emission is produced by X-ray irradiation and reprocessing in the disc.

However, at present, a detailed study based on simultaneous OIR and X-ray data of a single object from outburst to quiescence is still missing. Moreover, the population of BH systems as a whole is a factor of $\sim20$ optically brighter than the NS population at a given X-ray luminosity, both in outburst and quiescence \citep[see Fig. 2 in][]{russell06}. For LMXBs in outburst, \cite{russell06} and \cite{russell07} considered different possibilities, including the size of the accretion disc, however, the differences that arise when comparing BH and NS systems could not be fully explained.

Only simultaneous multi-frequency campaigns can help in disentangling the competing emission mechanisms. Here we present and compare for the first time the optical -- X-ray correlation of a NS system and a BH system over 6 orders of magnitude in X-ray luminosity, from outburst to quiescence. 
We selected two transient systems, the BH V404 Cyg (also known as GS $2023+338$) and the NS Cen X-4 (also known as V822 Cen). They are the only two transients with simultaneous X-ray and optical data both in outburst and quiescence. For a summary of the main binary system parameters see Tab. \ref{tab:source}.

On 15 June 2015 V404 Cyg entered a new activity period \citep{barthelmy15,barthelmy15b}. 
The new outburst started between June 2nd and June 8th, 2015, when a faint optical precursor was detected \citep{bernardini15}, and it was probably generated by a viscous-thermal instability triggered close to the inner edge of the truncated accretion disk \citep{bernardini16}. Near the outburst peak, optical variability was detected on timescale of tens of minutes \citep{marti16}. Sub-seconds timescale flares were also detected and presumably produced by an optically-thin synchrotron jet \citep{gandhi16}.
The data of the new outburst, which are not analyzed here, can be compared with the results of this work.
 
\begin{table*}
\caption{Main binary characteristics. The corresponding references are show in brackets below each value.}
\begin{center}
\begin{tabular}{ccccccccccc}
\hline
Source    & type     & outburst year & distance &  P$_{orb}$ $^a$ & incl.           & companion    &  M$_{c}$ $^b$        &      M$_{p}$ $^c$       & N$_{H}$               & A$_{V}$ \\
          &          &               &  (kpc)   &   (days)        &  deg            &              &  (M$_{\odot}$)   &  (M$_{\odot}$)      & 10$^{21}$ (cm$^{-2}$) &         \\
\hline
Cen X-4   & NS$^d$   & 69, 71, 79    & 1.2      & $0.6290522(4)$  & $32\pm^{8}_{2}$  &  K3-7 V     &  $0.23\pm0.10$   & 1.4  $^e$           & $0.80\pm0.08$                 & $0.31\pm0.16$  \\
          &          &   [1,2,3]     & [2,4,5]  &  [6]            &    [7]           &   [4,8]     &      [6]         & [6,7]               & [9]                   & [10]\\
\hline
V404 Cyg & BH & 38, 56, 89, 15 $^f$  & $2.39\pm0.14$  & $6.4714(1)$ &  $67\pm^{3}_{1}$       & K0($\pm1$) III-V  & $0.7\pm^{0.3}_{0.2}$  & $9\pm^{0.2}_{0.6}$  & $12\pm4$ $^g$  & $4.0\pm0.4$  \\
         &    &    [11,12,13]        &    [14]        & [15]        &   [16]          & [17,18,19]        & [20] & [21] & [22,23] & [18,19] \\
\hline
\end{tabular}
\label{tab:source}
\end{center}
\begin{flushleft}
$^a$ The spin period of Cen X-4 is currently unknown. \\
$^b$ Companion mass. \\
$^c$ Primary mass. \\
$^d$ The detection of an X-ray type I burst during the decay of the 1979 outburst, unambiguously showed that this binary hosts a NS \citep[]{matsuoka80}. \\
$^e$ This value is in between that measured by \cite{casares07}, $1.14\pm0.45$ M$_{\odot}$, and by \cite{shahbaz14}, $1.94\pm^{0.37}_{0.85}$ M$_{\odot}$. \\
$^f$ During June 2015, the source entered a new outburst state \citep{barthelmy15}, shortly after followed by a second activity period on December 2015 \citep{barthelmy15b}.\\
$^g$ The uncertainty corresponds to the standard deviation of the values of N$_{H}$ of observations 3--7 reported in \cite{zycki99b} together with that of \cite{rana15}. \\

References: [1] \cite{conner1969}; 
[2] \cite{kuulkers09}; 
[3] \cite{kaluzienski80};
[4] \cite{chevalier89}; 
[5] \cite{gonzalez05}; 
[6] \cite{casares07};
[7] \cite{shahbaz14};
[8] \cite{davanzo05};
[9] \cite{bernardini13}; 
[10] \cite{russell06}; 
[11] \cite{makino89}; 
[12] \cite{richter89}; 
[13] \cite{zycki99}; 
[14] \cite{millerjones09}; 
[15] \cite{casares94}; 
[16] \cite{khargharia10}; 
[17] \cite{wagner92}; 
[18] \cite{casares93}; 
[19] \cite{hynes09}; 
[20] \cite{shahbaz94}; 
[21] \cite{khargharia10}; 
[22] \cite{zycki99b}; 
[23] \cite{rana15}. \\
\end{flushleft}
\end{table*}

\section{Observations}
\label{sec:obs}

We searched the literature for quasi-simultaneous (within a day; $\Delta\,t\leq1$ d) X-ray and optical observations, which were carried out using the same optical filter in outburst and quiescence. For the X-ray data, we selected the 3--9 keV energy range to match previous studies \cite[see e.g.][]{corbel08}. 
To convert from count rates and fluxes to luminosities we used the values of N$_{H}$, A$_{V}$, and distance (d) reported in Tab. \ref{tab:source}. Moreover, for the outburst X-ray data of both Cen X-4 and V404 Cyg in the hard state, we assume a typical power law spectral shape, with spectral index $\Gamma=1.6\pm0.1$, as normally used for hard state sources \citep[e.g.][]{russell06}.

For Cen X-4 we used the 1979 outburst data as reported in \cite{russell06}, where $\Delta\,t\simeq0.5$ d. Those are V-band optical monochromatic luminosities, together with 2--10 keV luminosities that we converted to 3--9 keV luminosities using \textsc{WebPIMMS}\footnote{https://heasarc.gsfc.nasa.gov/cgi-bin/Tools/w3pimms/w3pimms.pl}. We subtracted the companion star contribution by assuming for it m$_{V}=18.4\pm0.1$ \citep[dereddened absolute magnitude;][]{chevalier89}, which translates to L$_{V}=1.55\pm0.14\times10^{32}$ erg/s.
Quiescent Cen X-4 data are from \cite{bernardini13}, who used strictly simultaneous optical-UV and X-ray \Swift\ data collected between May and August 2012. We converted the V-band count rate to monochromatic luminosity. We first dereddened the optical
data and then, as for the outburst data, subtracted the
companion star contribution. We converted the quiescent 0.3--10 keV count rate to 3--9 keV luminosity using the average spectrum presented in 
\cite{bernardini13} and the best-fitting spectral model presented in \cite{chakra14}, which consists of a NS atmosphere 
component plus a cutoff-powerlaw with $\Gamma=1.02\pm0.10$ and $E_{cutoff}=10.4\pm1.4$ keV). We found that 1 c/s (0.3--10 keV) corresponds to L$_{3-9\rm\,keV}=1.28\pm0.04\times10^{33}$ erg/s.

For V404 Cyg in outburst (1989), we used the R-band luminosity from \cite{russell06} adopting the updated source distance and reddening. We also subtracted the companion star contribution (L$_{R}=2.53\pm0.44\times10^{34}$ erg/s) estimated from the fit to the SED (see Fig. \ref{fig:sed} and Sect. \ref{subsec:sed}). X-ray data are instead from \cite{corbel08}; we converted their 3--9 keV fluxes to luminosities. The source showed high amplitude variability in the first phase of the outburst, with a short transition to a thermal state \citep[see][and references therein]{corbel08}. Consequently, in order to use only data from the hard state where both $\gamma$ and N$_{H}$ are constant within uncertainty \citep[see observations 3--7 in Tab. 2 of][]{zycki99b}, we limited our analysis to the pointings performed after MJD=47685. 
The R-band and X-ray data were quasi-simultaneous. However, by fitting the R-band flux decay with an exponential function, which provides a very good description of the decay \citep[as expected in the DIM;][]{lasota01}, we were able to estimate the R-band luminosity at the exact time of each of the X-ray pointings. 
Quiescent V404 Cyg data are from \cite{hynes04} who used simultaneous optical and X-ray $Chandra$ (0.3--7 keV) data collected on 2003 July 28/29. Optical data are taken in two close bands ($6300-6500$ \AA\ plus $6620-6820$ \AA) that lie within the R-band filter. We converted flux densities to luminosities using the R-band central wavelength (6580 \AA). We dereddened using $A_{R}=A_{V}\times\,y=4\times0.751$, where $y$ is from \cite{cardelli89}. In analogy with \cite{hynes09} and \cite{zurita04}, we subtracted the contribution from the companion star by using the lower envelope of the variability to define an upper limit on it. We converted the quiescent X-ray count rate to 3--9 keV luminosity using $\Gamma=1.85$ \citep{rana15}.

In Fig. \ref{fig:corr} we plot optical (L$_{opt}$) vs X-ray (L$_{3-9 \rm keV}$) luminosity in units of $10^{36}$ erg/s for all data in the present study. Uncertainties are, hereafter, at $1\sigma$ confidence level. In addition, radio to optical/UV data were collected to construct SEDs of Cen X-4 and V404 Cyg (see Fig. \ref{fig:sed} for data references). The IR-optical-UV data were dereddened using the law of \cite{cardelli89} as above. Data used for the SEDs were taken within a day, unless stated otherwise, by a range of MJD values in Fig. \ref{fig:corr} and Tab. \ref{tab:sed}; quiescence data were taken on several dates.

\section{Results}

\subsection{Optical -- X-ray correlations}

In order to assess whether an optical -- X-ray luminosity correlation is present, both in outburst and quiescence and find out whether the same correlation (e.g a single power law) extends across the two  states, we used the following procedure. We first fitted, separately, the quiescent and outburst data with a power law model, $y=ax^{\gamma}$, and we compared the two derived slopes ($\gamma$) and normalizations ($a$). Then, if these were found to be consistent with each other, we fitted together the whole data set, covering about six orders of magnitude in X-ray luminosity. To measure the statistical significance of the correlation, we used a Spearman's rank correlation test. We report the Spearman's correlation coefficient $\rho$, the null hypothesis probability $P$, $\gamma$, and $a$ in Tab. \ref{tab:corr}.

\begin{table}
\caption{Correlation fit results. O is Outburst (hard state), Q is quiescence and O-Q is the two data sets combined. The fit is made with a powerlaw of the form $y=ax^{\gamma}$, while $\rho$ is the Spearman's correlation coefficient.}
\begin{center}
\begin{tabular}{ccccc}
\hline
\multicolumn{5}{c}{Cen X-4} \\
\hline
State            & $\gamma$      & $a$                 & $\rho$  &        $P$ ($\sigma$) $^s$        \\
\hline
O       & $0.58\pm0.08$   & $0.0022\pm0.0003$   & 0.92   & $4.1\times10^{-6}$ (4.5)    \\
Q   & $0.50\pm0.06$   & $0.005\pm^{0.004}_{0.002}$   & 0.55   & $4.7\times10^{-6}$ (4.4)    \\
O-Q & $0.44\pm0.01$ & $0.0028\pm0.0002$   & 0.76   & $2.2\times10^{-15}$ ($>8$)  \\
\hline
\multicolumn{5}{c}{V404 Cyg} \\
\hline
O      & $0.56\pm0.03$   & $0.58\pm0.02$       & 0.91   & $3.6\times10^{-18}$ ($>8$)  \\
Q      & $0.46\pm0.01$ & $0.045\pm0.003$       & 0.68   & $3.8\times10^{-19}$ ($>8$)  \\
\hline
\end{tabular}
\label{tab:corr}
\end{center}
\begin{flushleft}
$^a$ In parenthesis we report the confidence level in $\sigma$.\\
\end{flushleft}
\end{table}

For Cen X-4 both in outburst and quiescence, the Spearman's rank test shows that the V-band and X-ray data are positively correlated. Moreover, the two power laws are consistent within $1\sigma$ uncertainty in $\Gamma$ and within $2\sigma$ in $a$. Consequently, we performed a single fit over the whole luminosity range. 
For V404 Cyg, in both cases, Spearman's rank test shows that the R-band and X-ray data are positively correlated.  However, while the two power law slopes are barely consistent within $3\sigma$, the two normalizations are inconsistent, the one from the outburst being much higher.
If we extrapolate the outburst best fitting power law downward to the bottom range of the explored X-ray luminosity (at $L_{X}=1.5\times10^{31}$ erg/s) we find that the measured quiescent optical luminosities are a factor of $\sim4.0$ fainter than expected. If we do the opposite and  extrapolate upward the best fitting quiescent power law to  $L_{X}=5.7\times10^{36}$ erg/s we find a factor of $\sim15$ difference. In Fig. \ref{fig:corr} we also show the best fitting powerlaw models.

We conclude that a single, positive correlation, with a slope $\gamma=0.44\pm0.01$, extending from outburst to quiescence represents well the Cen X-4 data. For V404 Cyg, instead, the slope of the correlation slightly decreases from outburst ($\gamma=0.56$) to quiescence ($\gamma=0.46$). However, the optical quiescent data are a factor of 4--15 fainter than expected, based on the outburst correlation alone. Moreover, when directly comparing V404 Cyg and Cen X-4, the optical luminosity of the BH correlation is a factor of about {\b 160--280} higher than that of the NS in outburst (the values of 160 and 280 are measured at $L_{X}=1\times10^{35}$ erg/s and $L_{X}=1\times10^{37}$, respectively). This factor is about 10 times smaller in quiescence.

\subsubsection{Systematic}

\begin{figure}
\begin{center}
\includegraphics[angle=0,width=3.65in]{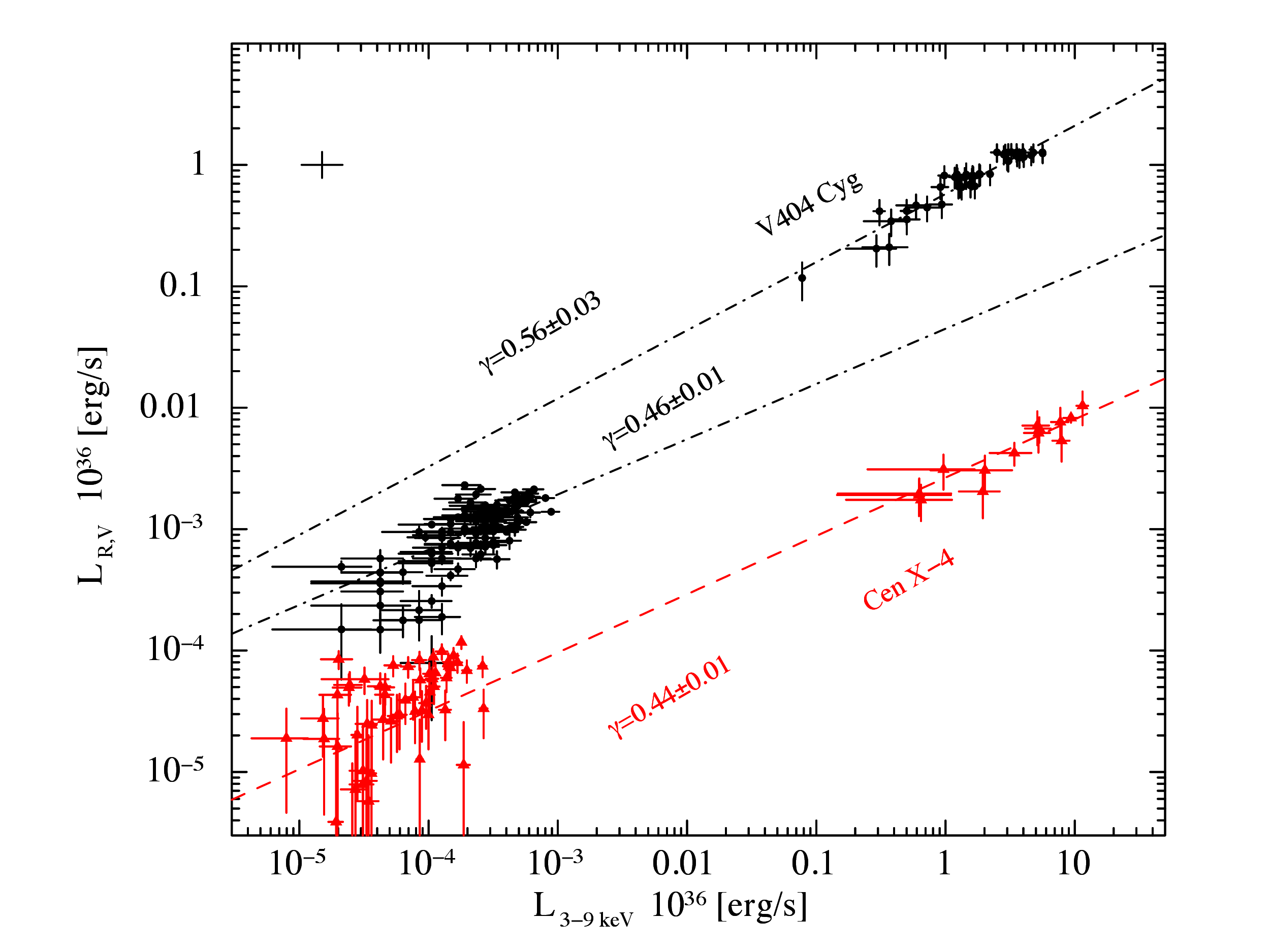}
\caption{Optical -- X-ray (3--9 keV) correlation. Black circles represent V404 Cyg data (R-band). Red triangles represents Cen X-4 data (V band). The black dot dashed lines is a power law fit to the V404 Cyg outburst and quiescent data, separately. The red dashed line is a power law fit over the whole Cen X-4 data set, from outburst to quiescence. The value of the power law index $\gamma$ is also reported. In the upper left corner we also show the average systematic uncertainty level, about a factor of 1.44 and 1.28 in optical and X-ray luminosity respectively (black cross).}
\label{fig:corr}
\end{center}
\end{figure}

In the top left corner of Fig. \ref{fig:corr} we show the average (between Cen X-4 and V404 Cyg outburst and quiescent data) systematic uncertainty, a factor of 1.44 and 1.28 in optical and X-ray luminosity, respectively. The X-ray systematic is the combination in quadrature of the uncertainties of the distance, the absorption, and the spectral shape (the uncertainty on the powerlaw photon index for V404 Cyg and on the conversion factor from 0.3--10 keV c/s to L$_{3-9\rm\,keV}$ for Cen X-4).
In the case of the optical data it includes the uncertainty in the distance, the extinction, the characterization of the companion star lower envelope ($\sim10\%$ for V404 Cyg), and a factor of $10\%$ systematic error in the measure of the quiescent flux (for V404 Cyg).
The total uncertainty in X-ray and optical luminosities is the combination in quadrature of the uncertainty of the X-ray and optical fluxes (plotted on each point) and the systematics. The effect of the systematics is not included in the correlation analysis for each source nor in the following sections, e.g. when fitting the corrected correlations.

\subsection{SEDs}
\label{subsec:sed}

The optical emission processes in outburst and quiescence can also be constrained by identifying components in the broadband spectrum. In Fig. \ref{fig:sed} we present the radio to optical/UV broadband spectra of V404 Cyg and Cen X--4 during outburst and quiescence taken from the literature. Displayed spectra are those in which multiple bands were observed on or around the same dates (the date ranges are given in the figure key).

\subsubsection{V404 Cyg}
The continuum optical emission of V404 Cyg close to the outburst peak (MJD 47676) was dominated by a blue component ($\alpha_{opt} > 0$) which is plausibly due to X-ray heating of the accretion disc, whereas during the decay (e.g. MJD 47728-9) the continuum was redder ($\alpha_{opt} < 0$) and consistent with a jet synchrotron spectrum \citep[see also][]{wagner91,russell13}. Both components are visible in the spectrum at different stages of the outburst decay. In the soft state (MJD 47676), the radio emission was bright and optically thin ($\alpha_{rad} < 0$; probably originating from discrete ejections) and its (power law) extrapolation cannot make a significant contribution to the optical emission. As the source faded and returned to the hard state, the radio spectrum flattened (MJD 47681--47683) then became inverted (MJD$\geq47696$), a behavior typical of the steady, compact jets commonly seen in the hard state \citep{han92,corbel08}. The maximum possible jet contribution to the optical emission, as estimated by extrapolating the radio power law, increased with time as the source evolved from the soft to the hard state. The flux of the irradiated disk component also decreased. In table \ref{tab:sed} the estimated star, disk and jet contributions are given for each date when these parameters can be constrained by the data. We report the value in the V-band, which is where most data are available (the derived percentages are consistent in the R-band).

In the following we describe in detail the way in which we estimate the companion star, disk and jet contribution in V404 Cyg for each date. The quiescent companion star is modeled as a blackbody at a temperature of 4570 K with a normalization that assumes $86.5\%$ of the quiescent $R$-band flux is stellar light \citep[see][and references therein]{hynes09}. The contribution of this blackbody to the $V$-band flux on each date is first estimated. The maximum jet contribution is the flux density of the extrapolated radio power law as a fraction of the observed flux density, with the quiescent companion contribution subtracted. The disk contribution is constrained from the remaining part of the observed flux density that cannot be accounted for by the star or jet. Before the transition to the hard state (MJD$<47683$), the jet made only a minimal contribution to the optical flux, $< 10\%$. The optical spectrum was also blue, and can be very crudely fitted by a single temperature blackbody (one example is shown; on MJD 47683 the spectrum can be described by a $\sim 11 500$ K blackbody), which most likely represents irradiation on the disk surface \citep[e.g.][]{hynes09}.

From MJD 47698 onwards, the jet could account for more than $50\%$ of the $V$-band flux density, and on the only date on which more than one optical/IR filter was acquired (MJD 47728--9), it is red and inconsistent with a blackbody. Instead, the fluxes on this date can be described by a jet spectrum with a break at infrared frequencies (Russell et al. 2013). Since the $V$-band flux density ($\sim 6\times10^{14}$ Hz) was considerably lower than the infrared ($\sim10^{14}$ Hz) flux densities, and the infrared to optical spectral index was $\alpha_{\rm thin} = -0.89 \pm 0.11$, the disk contribution was plausibly to be low at this time. This spectral index represents the slope of the optically thin synchrotron emission near the jet base if there is no disk contribution.
As a further check, we explored a wide range of possible disk contributions ($0\%-40\%$) and see how the jet spectral index would change accordingly. Then, we rejected those jet spectral index values that, also based on the comparison with other BHXBs, are unlikely (see below). 
In Fig. \ref{fig:alpha} we show the spectral index $\alpha_{\rm thin}$ of the jet spectrum as a function of the $V$-band contribution of the irradiated disk (the low-level contribution of the star has been subtracted from each waveband before fitting). 
If the disk contributed more than $\sim 10\%$ of the $V$-band flux, the jet spectral index would be steeper, $\alpha_{\rm thin} < -1.0$;  a disk contribution of $40\%$ would translate to $\alpha_{\rm thin} = -1.5 \pm 0.2$. Such a steep index is unusual, but not unheard of. A recent compilation of measurements of $\alpha_{\rm thin}$ from BHXBs shows the spectral index is typically -0.7 -- -0.8, but can be as steep as -1.4 \citep{russell13}. However, all the sources with $\alpha_{\rm thin} \leq -1.0$ accrete at low luminosity \citep[$L_{\rm X} \leq 5 \times 10^{-3} L_{\rm Edd}$; see also][]{shahbaz13} and in one source $\alpha_{\rm thin}$ was seen to increase with luminosity, and the steeper indices could be explained by a contribution of thermal particles in the jet at low accretion rates \citep[][]{russell10,shahbaz13}. If this were the case, then we would not expect the spectral index of the jet in V404 Cyg to be steep at the time of this broadband spectrum because its luminosity was high, with $L_{\rm X} = 2 \times 10^{-2} L_{\rm Edd}$. If $\alpha_{\rm thin} \geq -1.0$ then the disk contributed less than $\sim 14\%$ of the $V$-band flux (Fig. \ref{fig:sed}), with the jet contributing $78\%-92\%$. We suspect that there must be some disk contribution, because we would not expect the disk to fade so rapidly compared to the jet, so we estimate the disk contribution to be $\sim10\pm4\%$ on MJD 47728--9.\\

In quiescence, the radio spectrum appears to be approximately flat on three dates \citep[see also][]{gallo05,hynes09}, but the normalisation changes ranging from 0.19 to 0.5 mJy, demonstrating jet variability in quiescence \citep[see also][]{hynes09}. Even higher radio flux densities up to 1.5 mJy have been reported \citep{hjellming00} and imply a radio flux range of a factor of $\sim 8$ (but we do not include it in Fig. \ref{fig:sed}, right panel, because no radio frequency was reported). There is an excess of mid-IR emission (in the range $10^{13}-10^{14}$ Hz) above the star blackbody, which could originate in the jet if the flat radio spectrum extends to mid-IR frequencies \citep[see Fig. \ref{fig:sed}, right panel and see also][]{gallo07,hynes09}. By extrapolating the three radio spectra to higher frequencies, we are able to estimate the maximum contribution of the jet to the quiescent optical emission, which must be at maximum $3-12\%$ in $V$-band depending on the choice of radio spectrum (in Table \ref{tab:sed}, the values for the middle of the three radio spectra are shown). The disk contributes up to $17\%$, with the star producing the majority of the quiescent flux \citep[see also][]{pavlenko96,shahbaz03,hynes04,hynes09,zurita04,gallo07,xie14}.

\subsubsection{Cen X-4}

\begin{figure*}
\begin{center}
\begin{tabular}{cc}
\includegraphics[angle=-90,width=3.48in]{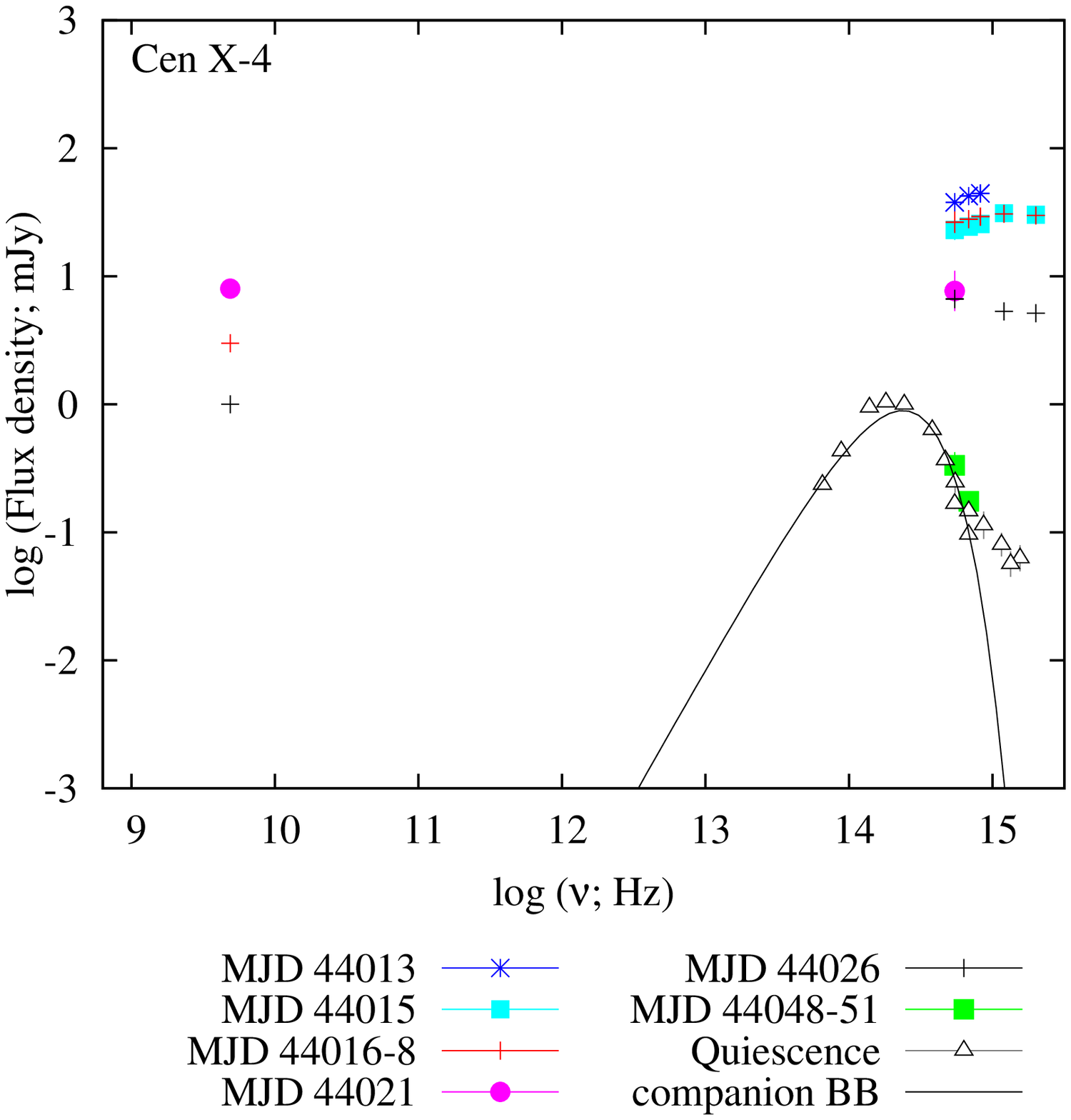} &
\includegraphics[angle=-90,width=3.38in]{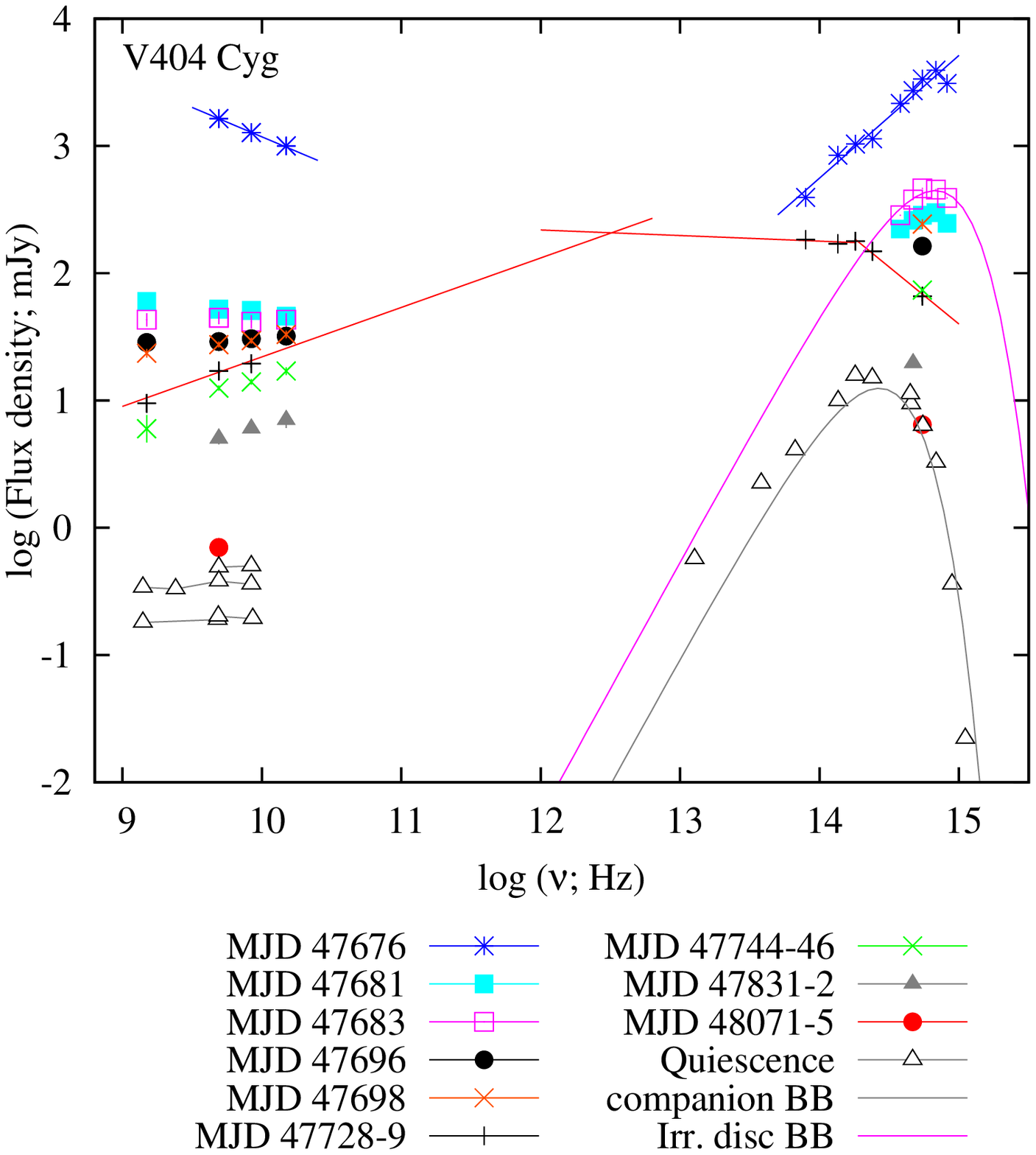} \\
\end{tabular}
\caption{ 
{\it Left panel}: Broadband, radio to optical (dereddened) spectra of Cen X-4 during its 1979 outburst and quiescence. The black line is a blackbody with a temperature of 4050 K representing the companion star.
Data are from: \cite{blair84,bernardini13,canizares80,hjellming79,kaluzienski80,shahbaz93,vanpar94,wang14}.
{\it Right panel}: Broadband, radio to optical (dereddened) spectra of V404 Cyg during its 1989 outburst and quiescence. The first spectra were obtained near the peak of the outburst and in general, time progresses from top to bottom. The red lines show the fit to the jet spectrum obtained on MJD 47728--9 \citep{russell13} and the grey and purple curves show single temperature blackbody approximations to the data in quiescence and on MJD 47683, respectively. The radio data are taken from \cite{han92,gallo05,hynes09}; the grey lines joining radio data points taken on the same dates. The infrared, optical and UV data are from \cite{johnson89,gehrz89,casares91,casares93,leibowitz91,udalski91,wagner91,han92,hynes09}.}
\label{fig:sed}
\end{center}
\end{figure*}

The SEDs of Cen X--4 are shown in the left panel of Fig. \ref{fig:sed}. During outburst, optical and UV fluxes were measured, but only three radio detections were reported, and no infrared data were taken. Near the outburst peak (MJD 44013--44018), the optical emission was dominated by a blue component ($\alpha > 0$), similar to V404 Cyg. This component appears to peak (in flux density) in the near-UV, which may be the peak of the irradiated disk component \citep[see][]{blair84}, also seen in some BHXBs \citep[e.g.][]{hynes05}. As the source faded during the outburst decline, the flux varied by a factor of $\sim 8$ in five days at radio frequencies (MJD 44021--44026), while the optical flux remained constant (within errors) during the same dates. This suggests that the jet emission is unlikely to dominate the optical emission, otherwise one might expect a variable optical flux of similar amplitude. The optical spectral index became slightly redder as the source faded (MJD 44026), possibly due to the peak of the irradiation blackbody shifting to lower frequencies (the spectral index is bluer than expected for optically thin synchrotron emission from the jet).

In quiescence, both the companion star and disk contribute to the IR--optical--UV SED \citep[see also][]{bernardini13,baglio14,wang14}. The IR SED can be well fit by a single temperature blackbody from the companion star, with an effective temperature of 4050 K \cite[][also shown in Fig. 2 of this paper]{baglio14}. The optical--UV flux excess above the companion correlates with the X-ray flux \citep[][]{cackett13,bernardini13} which can be explained by reprocessing. Stringent upper limits on the flux and polarisation of the jet at optical--IR frequencies have also been reported \cite[less than $10\%$][]{baglio14}.

We conclude that for Cen X--4 the jet have made a negligible contribution to the optical emission in outburst and quiescence.

% Note: quiescent data are:
% R-band: Hynes data: 9.37 +- 0.09 mJy at 4.55e14 Hz. Star model: 8.10425203098379 mJy (to get 13.5% non-star contribution)
% V-band: Hynes data: 6.35 +- 0.12 mJy at 5.50e14 Hz. Star model: 5.24470837220171 mJy
% J-band: Star model: 12.2910034689619 mJy
% H-band: Star model: 10.5847320680526 mJy

\begin{table*}
\begin{center}
\caption{Evolution of the irradiated disk, and jet contributions, before (I) and after (II) the subtraction of the contribution of the companion star,
to the optical ($V$-band) emission of V404 Cyg as estimated from the broadband spectra.}
\vspace{-2mm}
\begin{tabular}{llllllll}
\hline
MJD&Spectral&$V$-band flux&Irradiated disk (I) &Jet (I) &Star & Irradiated disk (II) & Jet (II) \\
    &state  &density (mJy)& (\%)               &(\%)    &(\%) & (\%)                & (\%) \\
\hline
47676     &  Soft           &3361            & $\geq 99.6$ & $< 0.2$  & 0.2         & $\geq 99.8$  & $< 0.2$ \\
47681     &  Soft to Hard   &285             & $\geq 92.8$ & $< 5.4$  & 1.8         & $\geq 94.5$  & $< 5.5$ \\
47683     &  Soft to Hard   &464             & $\geq 90.3$ & $< 8.6$  & 1.1         & $\geq 91.3$  & $< 8.7$ \\
47690     &  Hard           &324             & $\geq 66.2$ & $< 32.2$ & 1.6         & $\geq 67.3$  & $< 32.7$ \\
47696     &  Hard           &163             & $\geq 65.0$ & $< 31.8$ & 3.2         & $\geq 67.1$  & $< 32.9$ \\
47698     &  Hard           &243             & $\geq 37.5$ & $< 60.3$ & 2.2         & $\geq 38.3$  & $< 61.7$ \\
47711--13 &  Hard           &15.5            & --          & $< 66.2$ & 33.8        & -            & $< 100$ \\
47728--9  &  Hard           &$65.8 \pm 6.1$  & $\leq 14.0$ & $78.0-92.0^2$ & 8.0    & $0-15$       & $85-100$  \\
47744--6  &  Hard           &$73.9 \pm 1.7$  & --          & $\leq 92.9$ & 7.1      & -            & $\leq 100$ \\
47831--2  &  Hard           &$19.6 \pm 0.7^1$& --          & $\leq 58.7$ & 41.3     & -            & $\leq 100$ \\
48052     &  Hard           &$9.5 \pm 0.4$   & --          & $\leq 44.8$ & 55.2     & -            & $\leq 100$ \\
48071--5  &  Hard           &$\sim6.5$       & $\leq 19$   & $\leq 19$   & $\sim81$ & $\leq 100$   & $\leq 100$ \\
% Quiescence& &$6.35 \pm 0.12$ & 5.5--17.4 & $< 11.9$ & 82.6 \\ note: using brightest radio SED
Various    & Quiescence     &$6.35 \pm 0.12$ & 5.5--17.4 & $< 11.9$ $^3$ & 82.6         & $31.6-100$   & $<68.4$  \\ % note: using middle radio SED
% Quiescence& &$6.35 \pm 0.12$ & 14.4--17.4 & $< 3.0$ & 82.6 \\ note: using faintest radio SED
\hline
\end{tabular}
\normalsize
\label{tab:sed}
\end{center}
\begin{flushleft}
All flux densities are dereddened. The soft to hard state transition occurred around MJD 47680--5. \\
$^1$ $R$-band flux densities (no $V$-band available). \\
$^2$Assumes the maximum disk contribution is $13.4\%$ of the observed $V$-band flux, which is implied by the inferred slope of the optically thin synchrotron jet emission (see text).\\
$^3$ The jet contribution is $<11.9\%$, $<5.9\%$, and $<3.0\%$ using the brightest, central, and faintest radio SEDs, respectively.
\end{flushleft}
\end{table*}

\begin{figure}
\begin{center}
\includegraphics[angle=-90,width=3.5in]{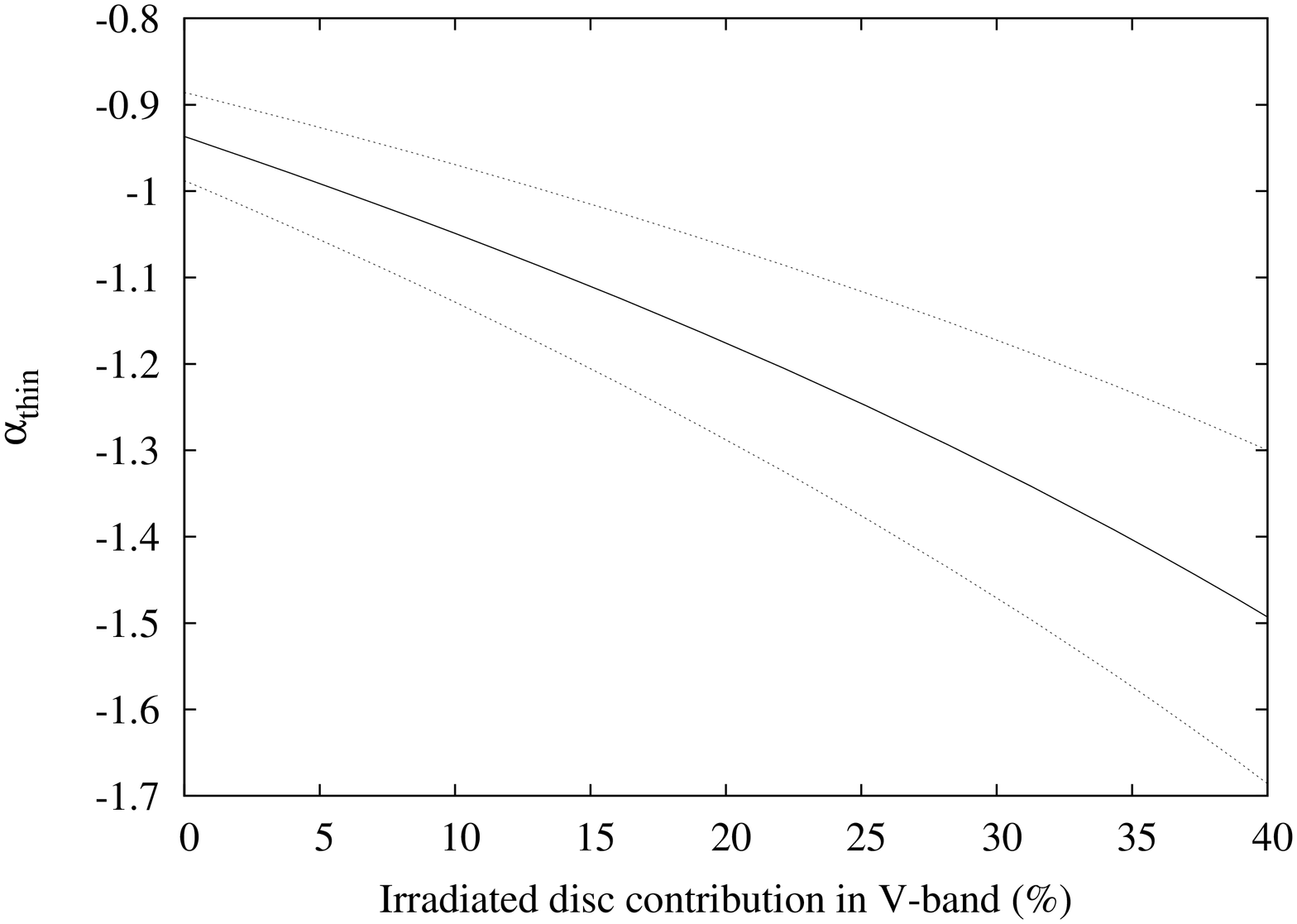}
\caption{The spectral index $\alpha_{\rm thin}$ of the optically thin synchrotron emission from the jet at optical--IR frequencies on MJD 47728--9 as a function of the contribution of the irradiated disk in the $V$-band. If the disk contribution exceeds $\sim 14\%$ of the flux, $\alpha_{\rm thin} < -1.0$, which is steeper than expected at this luminosity.}
\label{fig:alpha}
\end{center}
\end{figure}

\section{Discussion}
\label{sec:disc}

During outburst, we found that the optical luminosity of the BH system V404 Cyg is a factor of about 160--280 higher than that of the NS system Cen X-4, for a given X-ray luminosity. During quiescence, this factor is 10 times smaller. Moreover, we found that the normalization of the correlation of V404 Cyg changes between outburst and quiescence (see Fig. \ref{fig:corr}). 

We now explore what physical mechanisms can produce a higher optical flux in the BH system and the change in the slope and normalization of its $L_{X}-L_{opt}$ correlation. In particular, our goal is to test whether the $L_{X}-L_{opt}$ correlations of V404 Cyg and Cen X-4 overlap, once the main  known differences between the two systems are taken into account. 
To this aim we discuss and apply one by one the relevant corrections to the observed correlation  adopting Cen X-4 data as a benchmark.

\subsection{Corrections to the optical -- X-ray correlation}

\subsubsection{Bolometric X-ray emission}

The flux in the 3--9 keV range is just a fraction of the total X-ray flux ($F_{bol}$), for which we adopt the 0.5--200 keV range. $F_{3-9\,keV}=F_{0.5-200\,keV}\times\xi$, where $\xi$ is the bolometric correction factor. 
\cite{migliari06} provide the value of $\xi$ for BH and NS XBs in outburst, depending on the source type and spectral state, with the caveat that they derived $\xi$ by referring to the 2--10 keV flux (instead of 3--9 keV). Their values are $\xi=0.20\pm0.02$ for BH hard state, which includes the contribution of reflection \citep{nowak02} \footnote{To estimate the BH hard state conversion factor \cite{migliari06} used data from GX 339-4 presented in \cite{nowak02}, where reflection is clearly present. The reflection amplitude is correlated with the X-ray flux for GX 339-4 as it is for V404 Cyg \citep[see][]{zycki99b}.  \cite{nowak02} concluded that the correlation between X-ray flux and reflection is a general property of BH transients in the hard state.}, and $\xi=0.40\pm0.04$ for NS Atoll sources in the hard state. While Cen X-4 has not yet been properly classified \citep[see][for the atoll source definition and a catalog of LMXBs with their classification, respectively]{hasinger89,ritter03}, the detection of a type I burst \citep{matsuoka80} strongly suggests that Cen X-4 is a member of the Atoll class. Therefore, we assume that Cen X-4 is an Atoll source. 
In order to estimate the bolometric correction in quiescence we used, for Cen X-4, the $Swift$ X-ray data presented in \cite{bernardini13}. The $Swift$ observations, when summed together, provide the most extensive look at the source quiescence currently available ($\sim110$ ks). More recently, thanks to the high-energy sensitivity of the $NuSTAR$ observatory in the hard X-rays, \cite{chakra14} identified a spectral cutoff at $\sim10.4$ keV in the spectrum of Cen X-4 in quiescence. We applied their spectral model, consisting of a NS atmosphere plus a cutoff power law (spectral index 1.02), to the total $Swift$ X-ray spectrum. We extrapolated the model to the 0.5--200 keV range, compared the derived unabsorbed flux with that measured in the 3--9 keV range, and derived $\xi=0.16\pm0.02$. 

For V404 Cyg, we used the $Swift$ X-ray data ($\sim140$ ks) presented in \cite{bernardini14}, and a model consisting of a power law with spectral index 1.85 with a cutoff at 19 keV. This is what was recently found in quiescence using simultaneous \XMM\ and $NuSTAR$ by \cite{rana15}. We extrapolated the spectral model to the 0.5--200 keV range, compared the unabsorbed flux with that in the 3-9 keV range, and found $\xi=0.28\pm0.03$.

\subsubsection{X-ray reprocessing}

We found that for Cen X-4 a single power law correlation with a slope $\gamma$ close to 0.5 extends over six orders of magnitude in X-ray luminosity, suggesting that the same physical mechanism(s) operate(s) during outburst and quiescence. 
\cite{bernardini13} showed that during quiescence the X-ray and optical-UV fluxes are strongly correlated on timescale probably\footnote{The real structure of the quiescent disk is unknown and consequently the viscous timescale is hard to constrain.} shorter than the viscous timescale \citep[see also][]{cackett13}.
This suggests that the optical (and UV) -- X-ray correlation in quiescence is due to reprocessing and not to thermal instabilities propagating through the disk. The slope of the correlation can also provide constraints on the underlying emission mechanisms. \cite{vanpar94} showed that in the case of disk reprocessing the optical luminosity scales with the X-ray luminosity following the relation $L_{V}\propto L^{\gamma}_{X} R$, where $R$ is the accretion disk outer radius and $\gamma=0.5$ \citep[see also][for the way $\gamma$ changes in different frequency bands]{shahbaz15}.  
For Cen X-4, $\gamma$ is very close to 0.5 suggesting that X-ray reprocessing in the disk may be responsible for the optical luminosity. Consequently, we start assuming that X-ray reprocessing from the disk has a relevant role in the correlation and see whether this effect can account for the difference in optical luminosity between Cen X-4 and V404 Cyg. We stress that within uncertainty also the slope of the correlation of the BH V404 Cyg, both in outburst and quiescence is consistent with X-ray reprocessing.
From \cite{vanpar94}, by using Kepler's third law, we have $L_{opt} \propto L_{X}^{1/2}(M_{p}+M_{c})^{1/3}P^{2/3}$, where $M_{p}$ and $M_{c}$ are the primary (compact object) and companion mass, respectively. We find $L_{opt}{\rm (V404 Cyg)}/L_{opt}{\rm (Cen X-4)}\sim8-9$ \citep[the range mainly depends on the NS mass estimate that is 1.14 and 1.94 according to][respectively]{casares07,shahbaz14}. 
Once the V404 Cyg data are corrected for this effect (we used 8.5 that corresponds to a NS of 1.4 M$_{\odot}$), and the X-ray bolometric emission is considered, the two sets of quiescent data are much closer to each other, but there is still a factor $\sim20$ difference in optical luminosity during outburst (see Fig. \ref{fig:corr_corr}, left panel). 
We conclude that X-ray reprocessing alone cannot account for the difference in optical luminosity in outburst and that some further mechanism must make the BH system optically brighter. 

\subsubsection{Synchrotron jet emission}

In the following we try to identify which other mechanism can make V404 Cyg optically brighter with respect to Cen X-4. 
The SED of V404 Cyg in the hard state clearly reveals the presence of an optically thick synchrotron jet. Moreover, jet emission is plausibly detected at optical frequencies also in the flares of the 2015 outburst \citep[see e.g.][]{hynes15}. We showed that optical emission in the hard state is very likely jet-dominated starting from MJD 47711 (see Tab \ref{tab:sed}). From the SED where the jet break is found (MJD 47728--9), we estimate that the companion-subtracted jet contribution in the optical V-band is $85\%-100\%$. We assume that this range of values also applies to all the following hard state observations in the time range 47711--48075 MJD. The companion-subtracted jet contribution is instead much lower during quiescence, and the disk is probably dominating at optical frequencies ($31.6\%-100\%$). 

Cen X-4 was detected at radio frequencies ($\sim4.9\times10^{9}$ Hz) during outburst. The SED in the optical is fairly blue, likely due to a dominant disk contribution. There is no radio detection for Cen X-4 in quiescence, nor is there any evidence for an optical-IR  excess from the jet in Cen X-4 \citep[see also][]{baglio14,wang14}. Moreover, NS LMXBs presumably do not attain a jet-dominated state during quiescence \citep{gallo12}. 
The optical SEDs of Cen X-4 in outburst have a shape consistent with disk emission, a similar result to other NS systems in outburst. Any synchrotron emission would contribute significantly only at frequencies lower than the optical/IR bands in the case of Cen X-4 \citep[][]{maitra08,migliari10,lewis10}. We conclude that the jet contribution to the optical emission in outburst and quiescence for Cen X-4 is negligible. 
Summarizing, the comparison of the SED of V404 Cyg and Cen X-4 suggests that the jet is mainly responsible for shaping the difference between the two sources in outburst. 

In the following we test whether the slope of the correlation confirms these results. When comparing X-ray and optical (R-band) emission for the BH V404 Cyg in outburst, before applying any correction (see the uncorrected correlation in Fig. \ref{fig:corr}) we get $\gamma=0.56$. \cite{corbel08} found for V404 Cyg a power law correlation with a index $\gamma=0.51$, extending from outburst to quiescence, based on the comparison of simultaneous radio and X-ray data. They conclude that the correlation is produced by the emission from a compact jet. The value they found is fully consistent with what we found in outburst.
\cite{han92}, show that the spectral index of the self-absorbed, optically thick, synchrotron jet ($\alpha_{thick}$) in the radio band during the hard state decay of the 1989 outburst of V404 Cyg was between $\sim0.2$ and $\sim0.6$ \citep[see also][]{russell13}. We can use the 0.2--0.6 range to derive the expected theoretical relation between $L_{\nu,radio}$ (radio monochromatic luminosity at a given frequency) and $L_{X}$ for the case in which the jet is dominating the emission at these frequencies. Following the analytic model presented by \cite{heinz03}, assuming that the jet in the hard state is radiatively inefficient with $L_{X}\propto \dot{m}^{2}$, that the radiative efficiency is not changing with luminosity, and that the jet power is a constant function of the power released by accretion, we have $L_{\nu,jet}\propto\dot{m}^{(17/12-\frac{2}{3}\alpha_{thick})}$ \citep[see equation 7 in][]{russell13}. 
Rearranging for $L_{X}$, we have $L_{\nu,jet}\propto L_{X}^{0.51-0.64}$. Since the jet spectrum extends to optical/IR frequencies, and there is no strong dependency of the jet break frequency on luminosity \citep{russell13}, we expect this relation to hold also in the optical band ($L_{\nu,jet}\propto L_{opt}$). Summarizing, both from an observational \citep[see][]{corbel08} and theoretical \citep{heinz03} point of view, in the jet case we expect a power law relation between optical and X-ray luminosity with an index $\gamma$ very close to 0.5--0.6. This is indeed what we measure in outburst ($\gamma=0.56$), while, during quiescence, we measure a slightly lower value ($\gamma=0.46$). 

Starting from the uncorrected correlation, in order to remove the jet contribution in the outburst hard state, we consider that the jet produces $92\%$ of the optical emission (the mean of the estimated range when the companion contribution is subtracted). Once the jet contribution is removed, and the X-ray bolometric correction is applied, the outburst data of V404 Cyg appear to be closer to those from Cen X-4, but a difference of about one order of magnitude still remains. We notice that this corresponds to the difference expected if X-ray reprocessing in the disk takes place.

Let's now take a closer look at the two systems in quiescence. The X-ray spectrum of the BH V404 Cyg is well fitted by a simple power law model, while that of the NS Cen X-4 is more complex. In addition to the power law component there is also a thermal contribution arising from the NS surface. We estimate that the two components provide 50\% each of the 0.5--200 keV quiescent X-ray luminosity, in agreement with the results of \cite{bernardini13}. Since BH do not have a radiating solid surface, in order to properly compare V404 Cyg and Cen X-4 in quiescence, we need to also remove the X-ray luminosity that arises from the NS surface.

\begin{figure*}
\begin{center}
\begin{tabular}{cc}
\includegraphics[angle=0,width=3.60in]{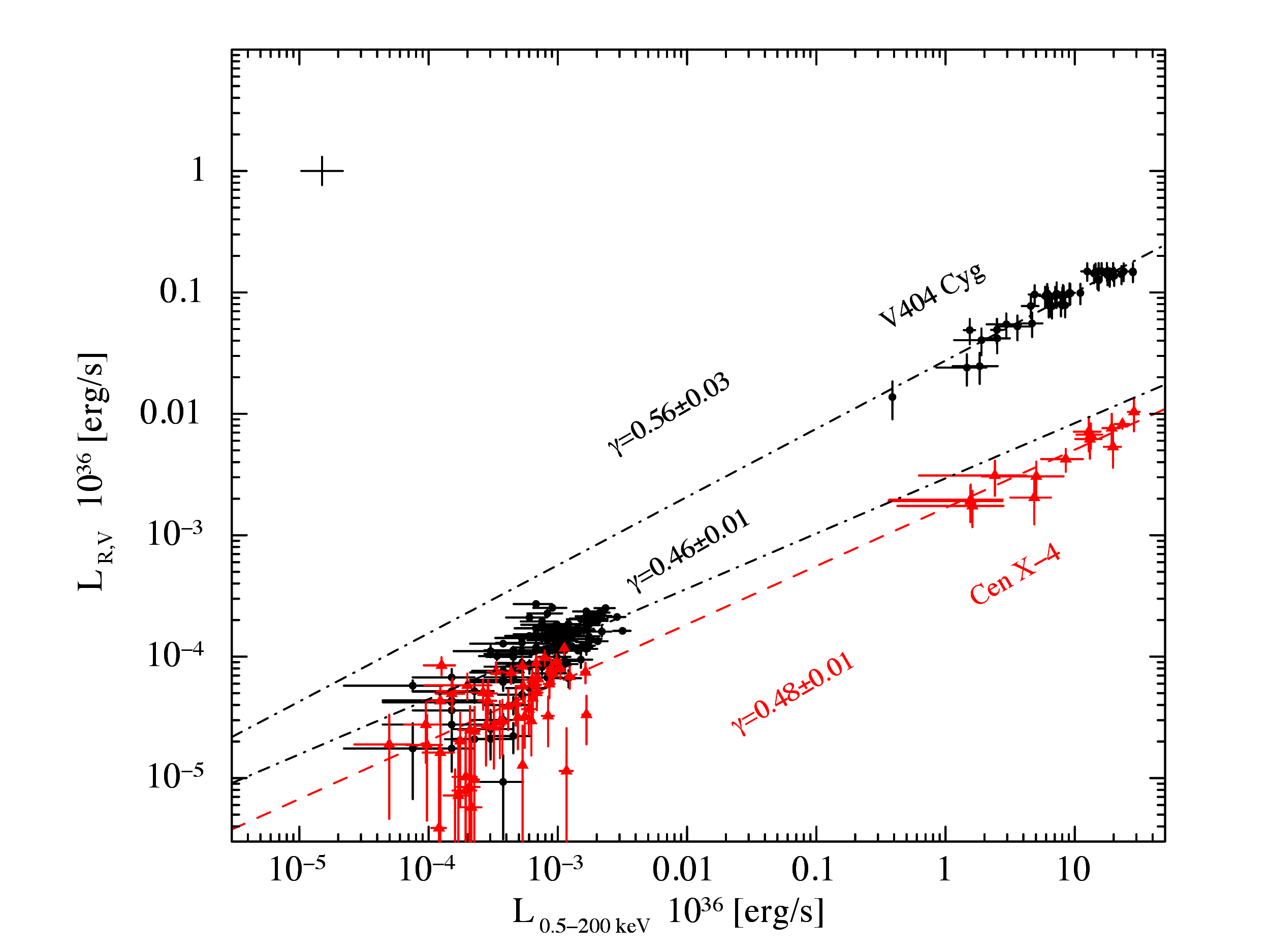} 
\includegraphics[angle=0,width=3.60in]{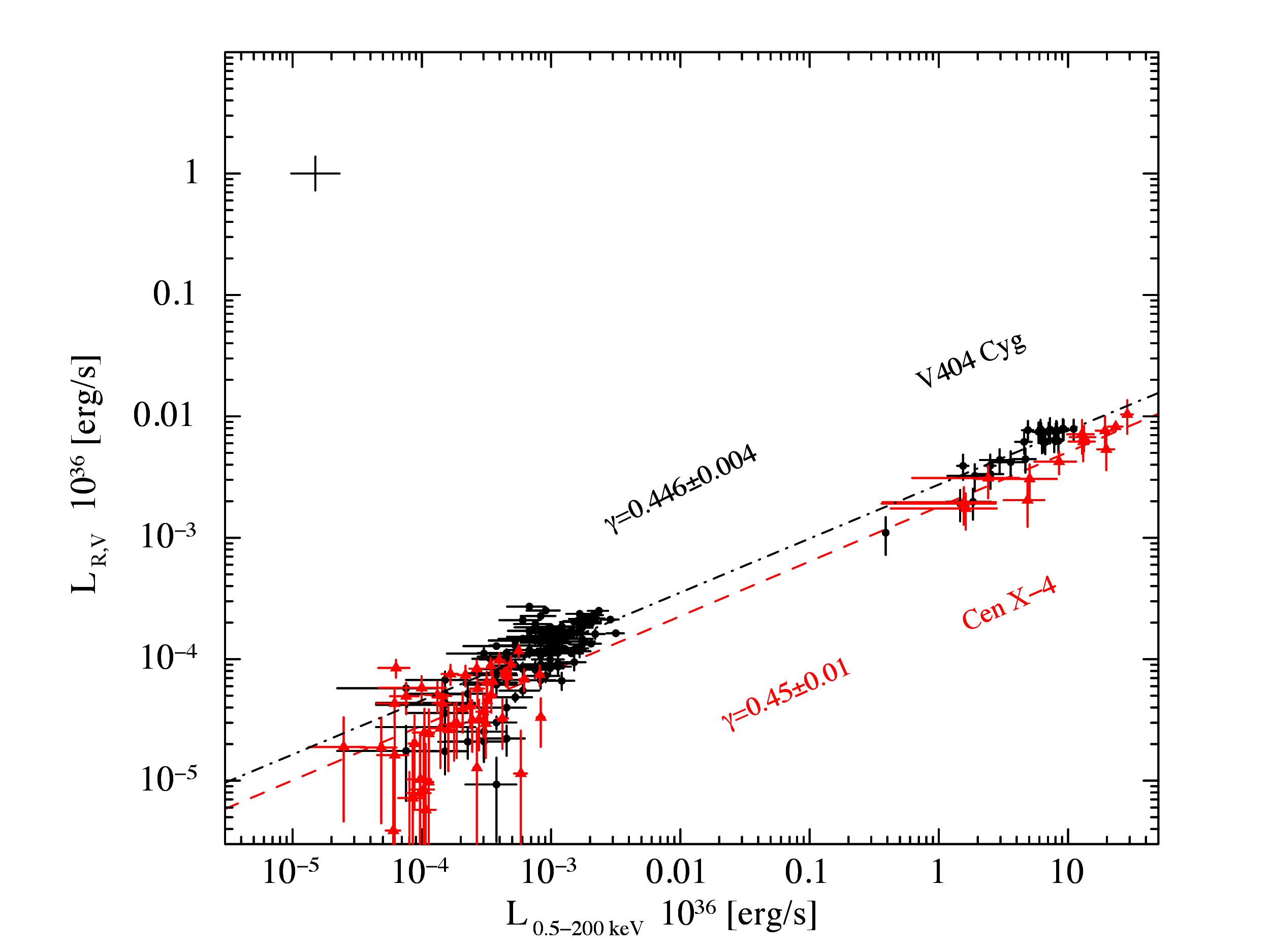}
\end{tabular}
\caption{Optical -- X-ray (0.5--200 keV) correlation. In both panels we used Cen X-4 as a benchmark. {\it Left panel:} The same as Fig. \ref{fig:corr}, but including the correction for orbital parameters (star masses and orbital period) and for the bolometric X-ray luminosity, which both affect the amount of optical emission we expect from reprocessing on the disk. Note that we have normalized the BH optical luminosity by a factor of 1/8.5 to account for the different disk sizes (see text for more details). The average systematic uncertainty level is a factor of 1.45 and 1.31 in optical and X-ray luminosity, respectively (black cross). 
{\it Right panel:} The same as the \textit{left} panel, but first subtracting the jet contribution for V404 Cyg in outburst ($92\%$ of the R-band luminosity once the companion star has been subtracted) and the contribution from the NS surface in quiescence for Cen X-4 ($50\%$ of the 0.5--200 keV luminosity), and then dividing by 8.5. For V404 Cyg in outburst hard state, we are only plotting the observations where the jet is dominating ($MJD\geq47711$) the emission at optical frequencies. The average systematic uncertainty level is a factor of 1.54 and 1.38 in optical and X-ray luminosity respectively (black cross).}
\label{fig:corr_corr}
\end{center}
\end{figure*}

We can look at the right panel of Fig. \ref{fig:corr_corr} (i.e. corrected correlation) as a comparison of the correlation
of V404 Cyg and Cen X-4 once the jet emission and the emission from
the NS surface have been removed, and assuming that the two systems have
the same disk size.
We also report in Tab. \ref{tab:correct} the intensity of all corrections we applied to generate this plot, starting from the raw data presented in Fig. \ref{fig:corr} (uncorrected correlation).
In more detail, for V404 Cyg we first account for the jet contribution in the hard state (subtracting 92\% the optical luminosity), which we have shown is the dominant emission process at this frequency starting from MJD 47711 (we are now only using observations from this date on). Then, we account for the difference in orbital parameters, namely the primary and secondary masses and orbital period (dividing the optical luminosity by 8.5) and for the bolometric X-ray correction. For Cen X-4 we also remove the blackbody-like contribution from the NS surface in quiescence (dividing the X-ray quiescent luminosity by 2), and keep only the power law spectral component.  
We notice that the outburst non-jet optical data and quiescent data of V404 Cyg now lie on the same correlation with $\gamma=0.446\pm0.004$. We further notice that Cen X-4 and V404 Cyg now display two correlations that perfectly match each other within $1\sigma$ uncertainty (Cen X-4 has $\gamma=0.45\pm0.01$).
We notice that both correlations have a slope close to 0.5, which is what is expected from X-ray reprocessing in the disk. It could be also consistent with a viscously heated disk case, where $L_{X}\propto L_{opt-UV}^{0.3-0.6}$ \citep[][]{shahbaz15}. However, we emphasize that in quiescence the optical emission of V404 Cyg \citep{hynes04} and the optical-UV emission of Cen X-4 \citep{cackett13,bernardini13} are correlated with that in the X-ray on short timescale (hundreds of seconds only for V404 Cyg), strongly suggesting that it is produced by X-ray reprocessing and not by an instability propagating in the disk on the viscous timescale, which is probably much longer than that.
 
We conclude that during outburst hard state of V404 Cyg, two main mechanisms, jet (dominant) and X-ray reprocessing, are responsible for the $L_{X}-L_{opt}$ correlation. During quiescence, instead, the jet contribution is minimal, and the X-ray reprocessing dominates. 
We ascribe to this the change in the normalization (and slope) in the uncorrected correlation (see Fig. \ref{fig:corr}). For Cen X-4, X-ray reprocessing is likely the dominating emission process both in outburst and quiescence. 
However, to confirm that disk reprocessing is taking place we also need to carefully compare the energetics (see Sect. \ref{sec:energetics}). 

\begin{table}
\caption{Summary of all corrections applied to the raw data in Fig. \ref{fig:corr} (uncorrected correlation) to get Fig \ref{fig:corr_corr}, right panel (corrected correlation). $\xi$ is the X-ray bolometric correction.}
\begin{center}
\begin{tabular}{ccccc}
\hline
\multicolumn{5}{c}{Quiescence} \\
\hline
Source         & $\xi$ $^a$ & jet $^b$ & disk size $^b$ & NS surface $^a$ \\
\hline
Cen X-4   & $\times6.3$ & -   & -      &  $\div2$  \\
V404 Cyg  & $\times3.6$   & -   & $\div8.5$   &  -     \\
\hline
\multicolumn{5}{c}{Outburst} \\
\hline
Source    & $\xi$ & jet& disk size & NS surface \\
\hline
Cen X-4   & $\times2.5$ & -      & -      & - \\
V404 Cyg  & $\times5$   & $\times0.08$  & $\div8.5$   & -   \\
\hline
\end{tabular}
\label{tab:correct}
\end{center}
\begin{flushleft}
$^a$ Correction applied to the X-ray data.\\ 
$^b$ Correction applied to the optical data.\\
\end{flushleft}
\end{table}

\subsubsection{Inclination and outer disk vertical size}

The amount of reprocessed light in the disk that reaches a far observer is proportional to $cos(i)$, where $i$ is the binary system inclination with respect to the line of sight. Taking the values from Tab. \ref{tab:source} ($i=32^{\circ}$ and $i=67^{\circ}$, for Cen X-4 and V404 Cyg respectively), we expect Cen X-4 to be a factor $\sim2$ optically brighter with respect to V404 Cyg for a given X-ray input due to this effect only. However, we point out that $i$ is model dependent and varies from paper to paper, e.g. \cite{shahbaz94} found $i=52\pm4^{\circ}$ for V404 Cyg. Moreover, the orbital inclination of Cen X-4 is not well constrained and within $\sim2.5\sigma$ it is consistent with $52^{\circ}$. We conclude that V404 Cyg correlation should be up-shifted by a factor of $\leq2$, depending on the exact value of its inclination.

The amount of reprocessed light from a geometrically thin irradiated disk, in the approximation of a X-ray point-like source located in its center in the disk plane, scales with $(H/R)_{ext}$, where $H$ and $R$ are the vertical size (semi-thickness) and the radius of the disk calculated at its outer edge, respectively \citep[see e.g. Sect. 5.4.4 in][for more details]{bernardini13}. We highlight that the geometry of the central X-ray source is not known, however, even if it is extended (e.g. a Comptonizing corona), it would remain almost point-like when seen by the outer disk.
In turn, $H/R$ is proportional to $\dot{M}^{3/20}$ and $R^{2/7}$ \citep[see equation 5.49 in][]{fkr02}, where $\dot{M}=L_{X}/\eta\,c^{2}$, $\eta\sim0.1$ is the efficiency, and the outer radius is of the order of $10^{11}$ cm and $10^{12}$ cm for Cen X-4 and V404 Cyg, respectively.
We explored how $H/R$ changes with the X-ray luminosity for V404 Cyg and Cen X-4 from outburst to quiescence and found that for a given X-ray luminosity Cen X-4 should be only a factor of 1.1--1.2 optically brighter. Moreover, we also found that in quiescence, for reasonable values of the viscosity parameter (e.g. $0.0001<\alpha<1$), the maximum amount of reprocessed luminosity is $\lesssim2\%$. 

We conclude that the effects due to the inclination and to the outer disk vertical size are minimal when comparing V404 Cyg with Cen X-4 in quiescence and outburst (hard state). Consequently, we do not further correct Fig. \ref{fig:corr_corr}, right panel for these effects. We stress that for other sources the inclination can be accounted for (when it is well constrained), but the disk semi-thickness is model dependent and largely unknown for most sources including V404 Cyg and Cen X-4 (this is especially true in quiescence, where the structure of the disk is still unclear).

\subsubsection{How to generate Fig. 4, right panel}

Here we summarize step by step, the method and equations we used to generate Fig. 4, right panel (corrected correlation), starting from the raw data (uncorrected correlation) and adopting Cen X-4 as a benchmark. This may be useful to update this figure in the future, including e.g. more data from other LMXBs or new outburst data from V404 Cyg and Cen X-4.

By transforming the observed correlations, we now aim at sketching a "standard" correlation that would hold for disk reprocessing only 
for LMXBs containing a BH or a NS, with: I - a given bolometric luminosity; II - emission from the jet removed; III - a given accretion disk size (which in turn implies a given primary mass, secondary mass, and orbital period); IV - emission from the NS surface removed; V - a given inclination with respect to the line of sight.

I  - The bolometric X-ray luminosity should be calculated using $F_{3-9\, keV}=F_{0.5-200\,keV}\times\xi$, where $\xi$ is 0.4 for NS systems in outburst and 0.2 for BH systems in outburst hard state. In quiescence we used 0.16 and 0.28 for Cen X-4 and V404 Cyg respectively. $\xi$ should be estimated for each system in quiescence when possible.  

II - The jet emission contribution at optical frequency should be removed (both in outburst and quiescence, if present). For V404 Cyg in outburst hard state it is 92\% of the companion-subtracted optical flux.

III - The jet-subtracted optical emission should be normalized to that of Cen X-4 using $L_{opt} \propto L_{X}^{1/2}(M_{p}+M_{c})^{1/3}P^{2/3}$. For Cen X-4 we adopt $M_{p}=1.4$ $M_{\odot}$, M$_{c}=0.23$ M$_{\odot}$ (mass ratio $q=0.16$), and $P_{orb}=0.6290522$ days.

IV - The NS surface X-ray emission must be removed (if present). It is $\sim50\%$ of the bolometric X-ray luminosity for Cen X-4.

V - The inclination should be normalized to that of Cen X-4 using $L_{opt} \propto cos(i)$, where $i=32^{\circ}\pm^{8^{\circ}}_{2^{\circ}}$ . We did not correct the figure for this effect because the measure of $i$ is model dependent and the inclination of Cen X-4 within uncertainty is consistent with that of V404 Cyg \citep[$i=52^{\circ}\pm4^{\circ}$,][]{shahbaz94}.

\subsection{Energetics}
\label{sec:energetics}

To unambiguously assess whether reprocessing is a valid explanation for the observed correlation, we must take into account that in the reprocessing scenario, the optical reprocessed luminosity is just a small fraction of the X-ray luminosity (a few per cent in outburst, less than $2\%$ in quiescence) if the disk is geometrically thin \citep{shakura73}. 

We start exploring the outburst data. In the case of the BH system we first subtract the dominating jet contribution in outburst. Moreover, we only account for the X-ray bolometric correction, and not for the accretion disk size. Then we compare the optical luminosity to that in the X-ray. From the power law fit to the correlation we measure $L_{V}/L_{X}\sim0.04-0.1\%$ for Cen X-4 and $L_{R}/L_{X}\sim0.6-2.4\%$ for V404 Cyg, two values fully consistent with the reprocessing scenario. Even considering a wider optical-UV range (e.g. not only the R or the V band), the ratio would remain small for both systems and consistent with reprocessing.

On the contrary, during quiescence the X-ray flux of both sources is much fainter. Consequently, in order to properly compare the energetics, we need to consider the contribution of the extra light (e.g. in excess of that coming from the companion) in the whole optical and UV band. 
For Cen X-4 we used the \Swift\ optical-UV and X-ray data (obsid$=0003532019$) presented in \cite{bernardini13}. From the SED we measured a total excess optical-UV luminosity (excluding the \Swift\ UVM2 band because it overlaps with others) of $2.3-3.8\times10^{32}$ erg/s that we compare with the X-ray luminosity. We here use the 0.01--200 keV luminosity, where we also include the extreme UV and the soft X-ray bands that are expected to produce reprocessing, with the caveat that the real spectral shape below 0.5 keV is not known. We extrapolate the X-ray spectral model, using the $1\sigma$ uncertainty on the parameters of the NS atmosphere and the broken power law. For the latter we used $\Gamma=1.02\pm0.1$ and $E_{cutoff}=10.4\pm1.4$ keV \citep[see][]{chakra14}. We estimate $L_{0.01-200\,keV}=1.3-1.7\times10^{33}$ erg/s. Correspondingly, we get a ratio $L_{opt-UV}/L_{X}\sim18-22\%$ that appears high to be only due to X-ray reprocessing in a geometrically thin disk. The ratio is higher than that previously reported by \cite{bernardini13} who used instead a simple power law spectra shape (e.g. without a energy cut off).

For V404 Cyg, unlike Cen X-4, the companion star dominates at all frequencies in the average SED and in particular in the UV band (see Fig. \ref{fig:sed}), so the disk contribution in the optical is likely small \citep[see also Fig. 1 in][]{hynes09}. 
Consequently, as we did for outburst data, we measured the excess emission in V404 Cyg in the optical only, and directly from the power law fit to the correlation, where we only account for the X-ray bolometric correction. We get $L^{min}_{R}\sim1.8\times10^{32}$ erg/s and $L^{max}_{R}\sim1.8\times10^{33}$ erg/s (where $L^{min}_{R}$ and $L^{max}_{R}$ are the minimum and maximum of the quiescent R-band excess). We estimate the X-ray luminosity using the high and low count rate spectra presented in \cite{bernardini14}, fitting it with a cutoff power law, where $\Gamma=1.85\pm0.15$ and $E_{cutoff}=19\pm^{19}_{7}$ keV \citep{rana15}. From the extrapolation of the X-ray spectral model we get $L_{0.01-200\,keV}^{min}\sim1.0\times10^{33}$ erg/s and $L_{0.01-200\,keV}^{max}\sim3.1\times10^{33}$ erg/s. Consequently, we measure $L_{opt-UV}/L_{X}\sim18-58\%$. We notice that this range is very sensitive to the power law parameters. Taking into account the $3\sigma$ uncertainty on it, we get $L_{opt-UV}/L_{X}\sim7-20\%$. 

We conclude that reprocessing in a geometrically thin disk can easily explain the optical-UV excess in outburst for both Cen X-4 and V404 Cyg (once the jet contribution is removed for the latter source). This matches the results of \cite{russell06} and \cite{russell07} for the global population of BH and NS systems. 
In quiescence, the amount of optical-UV light exceeds that expected from  reprocessing in a geometrically thin disk even if we include the contribution from the irradiated surface of the companion. The latter can only reprocess a few per cent of the X-ray emission \citep[e.g. $\leq4\%$ in the case of Cen X-4;][]{bernardini13}.
However, in the case of V404 Cyg, some residual intrinsic (e.g. not reprocessed) contribution from the disk could still be present, reducing the ratio $L_{opt-UV}/L_{X}$. Unfortunately, it is difficult to constrain the exact amount of this residual contribution with the currently available SED. 
We point out that in quiescence the real structure of the disk is still unknown. If the outer edge of the disk is thicker (e.g. because of the matter that accumulates on the stream impact point), or if the disk is warped, or if the X-ray source is not point-like, and/or the emission originates above the disk plane, like in the so called "lamp-post" geometry \citep{nayakshin01}, the maximum amount of reprocessed light would increase and could match the observed ratio. 
In this respect, we notice that \cite{dubus99}, using a self-consistent X-ray irradiated accretion disk model, showed that the outer disk of LMXBs can only be irradiated by the central X-ray source if the latter is located above the disk plane or if the disk is warped, or both (because the self screening from the inner disk). 
Moreover, we emphasize that the real spectral shape of both sources below 0.5 keV is not known. For example, the boundary layer of the disk could produce seed photons for reprocessing and its emission could peak below 0.5 keV. This would further reduce the ratio $L_{opt-UV}/L_{X}$.

\cite{cam&stell} suggested that for NS LMXBs the UV and X-ray emission could be produced at the shock between the matter transferred from the companion and the pulsar relativistic wind under the hypothesis that the NS posses weak, rapidly rotating, magnetosphere. In this case, accretion down to the NS surface maybe halted and a correlation between the UV and the X-ray power law flux is expected. However, for Cen X-4 the UV emission is also correlated with the thermal emission (which is changing in tandem with the power law one) suggesting that matter finally reaches the NS surface \citep{bernardini13}.

\section{Conclusions}
\label{sec:conc}

We have collected for the first time quasi-simultaneous optical (V, R bands) and X-ray luminosity, together with radio to UV SEDs, from outburst to quiescence, for two of the best studied transient LMXBs: the NS system Cen X-4 and the BH system V404 Cyg. 

We found that for Cen X-4 a strong correlation of the form $L_{opt}\propto L_{X}^{0.44}$ holds over 6 order of magnitude in X-ray luminosity. A similar correlation is found for V404 Cyg in outburst during the hard state, where $L_{opt}\propto L_{X}^{0.56}$ and in quiescence where $L_{opt}\propto L_{X}^{0.46}$. 
However, the optical quiescent data of V404 Cyg are a factor of 4--15 under-luminous, based on the extrapolation of the outburst correlation alone, namely, there is a change also in the normalization of the correlation before entering in the quiescent state. Moreover, we found the BH to be optically brighter than the NS at a given X-ray luminosity by a factor of 160--280 in outburst and a factor about 13-25 in quiescence. 

We found from the SEDs of V404 Cyg in outburst that the jet contributes the majority of the optical flux during the hard state, whereas in the soft state the (probably irradiated) disk dominates the optical emission. We identified the main physical mechanisms that make V404 Cyg optically brighter than Cen X-4 for a given X-ray luminosity. Once we consider the bolometric X-ray emission and we account for the fact that the optical emission during the BH outburst hard state is jet-dominated, while the jet very likely produces negligible contribution in the case of the NS (both in outburst and  quiescence), that the BH is more massive and has a larger accretion disk, and for Cen X-4 in quiescence we only consider the X-ray emission produced by the power law spectral component (namely the emission from the NS surface is subtracted), the two systems lie on the same correlation: a single power law with slope close to 0.5, extending from outburst to quiescence.

We have also shown that For V404 Cyg in outburst (hard state), two mechanisms, the dominant jet and X-ray reprocessing are responsible for the observed $L_{X}-L_{opt}$ correlation. During quiescence instead, the contribution of the jet is minimal. This is what probably produces the changes in the normalization and slope of the uncorrected correlation of V404 Cyg. For Cen X-4 in outburst, X-ray irradiation and reprocessing from the disk is very likely the dominating mechanism in producing the correlation. For both sources in quiescence (in particular for Cen X-4), the ratio between optical-UV and X-ray luminosity is above that expected from reprocessing in a geometrically thin accretion disk. We speculate that the emission from the boundary layer of the disk could peak below 0.5 keV, or that the outer disk could be thicker, the disk could be warped, and/or the irradiating X-ray source could be extended, and/or above the disk plane.

Based on the present study we suggest that the three main factors, the jet contribution, the accretion disk size (proportional to the orbital period), and the X-ray bolometric correction could account for the global difference between the optical/X-ray relationship of the population of BH and NS transients. The effects due to the system inclination should be also considered (but was marginal when comparing Cen X-4 and V404 Cyg). The average difference in optical luminosity for the two global populations is a factor of $\sim20$ \citep{russell06}, but it is expected to vary from case to case. This can be verified through detailed studies of the other systems.

In this respect, we generate a plot (Fig. \ref{fig:corr_corr}, right panel) where we show a {\it universal} $L_{opt}-L_{0.5-200\,keV}$ correlation for reprocessing only. The plot is normalized to the parameters of Cen X-4. The jet contribution in the optical band is removed, the primary mass is $1.4 M_{\odot}$, the mass ratio is $q=0.16$, and the orbital period is $P_{orb}=0.629$ days (which translates to a disk size $R_{out}\sim10^{11}$ cm), the X-ray emission from the NS surface is removed, and the orbital inclination is $i=32^{\circ}$. We provide details on how update this figure to include new data (from other sources and from new outbursts of V404 Cyg and Cen X-4).

\section*{Acknowledgments}

We acknowledge the anonymous referee for useful comments. 
FB acknowledges Koji Mukai for his precious suggestions and Tariq Shahbaz for the useful discussion. 
LS acknowledges partial support from ASI INAF I/004/11/1.
SC acknowledges the financial support from the CHAOS project ANR-12-BS05-0009 supported by the French Research National Agency. 

\bibliographystyle{mn2e}
\bibliography{biblio}

\vfill\eject
\end{document}